\documentclass[numberedappendix]{emulateapj}
\usepackage{amsmath}
\usepackage{rotating}
\usepackage{epsfig}

\def\simless{{\th \rlap{\raise 0.5ex\hbox{$\scriptstyle  {<}$}}
    {\lower 0.3ex\hbox{$\scriptstyle  {\sim}$}} \th }}  %< or of order
\def\simgreat{{\th \rlap{\raise 0.5ex\hbox{$\scriptstyle  {>}$}}
    {\lower 0.3ex\hbox{$\scriptstyle  {\sim}$}} \th }}  %> or of order
\def\greateq{{\th \rlap{\raise 0.5ex\hbox{$\scriptstyle  {>}$}}
    {\lower 0.3ex\hbox{$\scriptstyle  {-}$}} \th }}  %< or of order
\def\lesseq{{\th \rlap{\raise 0.5ex\hbox{$\scriptstyle  {<}$}}
    {\lower 0.3ex\hbox{$\scriptstyle  {-}$}} \th }}  %> or of order
\def\th{\thinspace}
\def\ts{{\raise 0.3ex\hbox{$\scriptstyle {\th \sim \th }$}}}

\newcommand{\etal}{\mbox{et al.}}
\newcommand{\saxjfull}{\mbox{SAX J1808.4$-$3658}}

\newcommand{\RXTE}{\textit{RXTE}}

\begin{document}

\shortauthors{Patruno \etal}
\shorttitle{Re-Flaring state, 1 Hz QPO and pulsations}

%\submitted{Submitted to ApJ Letters}
\title{1 Hz flaring in SAX J1808.4--3658: flow instabilities near the propeller stage}
\author{Alessandro Patruno \altaffilmark{1}, Anna
Watts\altaffilmark{1}, Marc Klein Wolt\altaffilmark{1}, Rudy
Wijnands\altaffilmark{1}, Michiel van der Klis\altaffilmark{1}}
\altaffiltext{1}{Astronomical Institute ``Anton Pannekoek,''
University of Amsterdam, Kruislaan 403, 1098 SJ Amsterdam,
Netherlands}

\email{a.patruno@uva.nl}

\begin{abstract}

We present a simultaneous periodic and aperiodic timing study of the
accreting millisecond X-ray pulsar SAX J1808.4-3658.  We analyze five
outbursts of the source and for the first time provide a full and
systematic investigation of the enigmatic phenomenon of the 1 Hz
flares observed during the final stages of some of the outbursts.  We
show that links between pulsations and 1 Hz flares might exist, 
and suggest they are related with hydrodynamic disk
instabilities that are triggered close to the disk-magnetosphere
boundary layer when the system is entering the propeller regime.

\end{abstract}

\section{Introduction}\label{1Hz:intro}

The low mass X-ray binary transient SAX J1808.4-3658 (hereafter J1808)
was discovered with {{\it Beppo}-SAX} in 1996 \citep{int98} and was
found to be an accreting millisecond X-ray pulsar
(AMXP, \citealt{wij98}) in 1998 when observed with the {\it Rossi X-ray
Timing Explorer} (\RXTE). Since then, four other outbursts have been
observed with \RXTE.  The X-ray lightcurves of the 5 outbursts under
consideration are remarkably similar in shape and duration. The
typical outburst duration is several weeks with a recurrence time of
$\sim 2.5$ yr; after 1998, outbursts occurred again in 2000, 2002,
2005 and 2008. The accretion rate increases steeply in the first 2--5
days of the outburst (\emph{fast rise}), then it stays relatively high
for a few days (\emph{peak}), reaching at most a few percent of the
Eddington rate.  After this, the X-ray flux has a \emph{slow decay}
lasting $\sim$ 10 days, before entering a \emph{fast decay} stage in
which the flux drops in 3--5 days.  The source then enters a low flux
state characterized by 3--5 day flares separated by intervals of
very low luminosity, the \emph{re-flaring} state that can last for
months, followed by \emph{quiescence}.

J1808 has shown 401 Hz pulsations during all the outbursts, at all the
luminosities observable by {{\it RXTE}} ($\simgreat
10^{34}\rm\,erg\,s^{-1}$), even during the re-flares \citep{har08,
har09}.  The re-flaring state was observed to last $\sim60$ days (MJD
$53550-53610$) in the 2005 outburst, followed by a low luminosity
state approximately 10 times brighter than quiescence that lasted for
another $\sim 60$ days~\citep{cam08}. In the 1998 outburst the {{\it
RXTE}} observations stopped immediately after the beginning of the
re-flares, while in the 2000 outburst, only the re-flaring state was
observed (for $\sim 100$ days, see \citealt{wij01}).  Thanks to the
good sensitivity of the {{\it XMM}-Newton} and {{\it Swift}-XRT}
satellite, \citet{wij03} and \citet{cam08} measured a minimum
luminosity of $\sim5\times10^{32}\rm\,erg\,s^{-1}$ between the flares
in the re-flaring state in the 2000 and 2005 outburst (assuming a
distance of 3.5 kpc). In 2002 \citep{wij04} and 2008 \citep{har09}
the re-flaring state was observed with {{\it RXTE}} for approximately 1
month.

\citet{cam08} interpreted the observed low luminosities as a signature
of the onset of the propeller regime. The propeller regime is
characterized by a Keplerian velocity in the innermost region of the
accretion disk that is slower than the rotational velocity of the
neutron star magnetosphere. In its original formulation \citep{ill75}
it was proposed to suppress the accretion flow onto the neutron star
surface. The gas, carrying part of the neutron star angular momentum,
was thought to be expelled and spin down the neutron star.
\citet{gho79a,gho79b} proposed that spin down and accretion
could occur simultaneously and recent MHD simulations
(\citealt{rom05}, \citealt{ust06}) show that two different propeller
regimes are possible: a strong propeller, characterized by a strong
outflow of gas, and a weak propeller, with no outflows.  In both cases
a magnetically channeled accretion flow onto the neutron star surface
is still expected consistent with the 401 Hz pulsations observed.  

\citet{wij04} reported a modulation at a repetition
frequency of $\sim 1$Hz that completely dominates the lightcurve
of J1808 in the 2000 and 2002 outbursts.  This $\sim1$ Hz modulation appears as
sudden intensification of the X-ray flux that are obvious in the
power spectra and sometimes are directly detected in the lightcurve.
A re-analysis of these data, along with 
a complete investigation of this phenomenon for the other
three outbursts (1998, 2005 and 2008), is presented in this paper. 

 The mechanism of the re-flares and the 1 Hz modulation is still
unclear, but it might be related to the onset of instabilities
expected for sources near the propeller stage (\citealt{ust06}).
It has been suggested that the fast decay and the re-flares are
related to cooling and re-heating fronts propagating through the disk
(\citealt{dub01}).  The heating fronts change the accretion disk
structure from a neutral to an ionized state, increasing the viscosity and
the mass transfer rate through the inner accretion disk.  If the
inner disk structure is influenced by this process, the
disk-magnetospheric boundary and/or the accretion process can be
modified as well, possibly producing hydrodynamic instabilities in the
accretion flow \citep{goo97, spr93, bil95}.

Whether the 1 Hz modulation is created by such instabilities is still
an open question. J1808 provides a unique opportunity to study this,
since it shows X-ray pulsations that can be observed simultaneously
with the aperiodic variability.  One of the reasons why the pulsations
can play an important role in understanding the mechanism behind the 1
Hz modulation is the pulse behavior observed during the 2002
outburst. A drift of $\sim 0.2$ cycles was observed in the pulse
phases of the fundamental frequency (but not in the first overtone),
starting and ending in coincidence with the beginning and the end of
the fast decay (\citealt{bur06}).  The pulse phase starts to drift in
coincidence with the beginning of the fast decay and ends when the
re-flares appear.  The interpretation of this drift is controversial.

\citet{bur06} concluded that the phase drifts appear
in coincidence with the onset of instabilities induced by accretion of
matter onto a weakly magnetized star, such as motions of the hot spot on
the neutron star surface. \citet{har08} concluded
that the observed phase drift might have been due to a motion of the
hot spot toward the magnetic pole as the inner accretion disk recedes
with decreasing luminosity (and thus decreasing mass accretion rate).

In this paper, we present the first comprehensive analysis of the 1 Hz 
modulation and its relation to the re-flares and 401 Hz pulsations 
in all outbursts of J1808. We discuss possible explanations for the 
onset of the 1 Hz modulation and possible reasons why it has been 
observed only in J1808 until now. We suggest the onset of 
accretion flow instabilities when J1808 enters the propeller stage
as the origin of the 1 Hz modulation.

\section{X-ray observations and data reduction}

\subsection{{{\it RXTE}} observations}
We reduced all the pointed observations with the {\it{RXTE}} satellite's
Proportional Counter Array (PCA, Jahoda et al. 2006)
that cover the outbursts of J1808.  

The aperiodic timing analysis was done using GoodXenon data with a
time resolution of $2^{-20}$s and Event data with a time resolution of
$2^{-13}$s. The data were binned into
1/8192 s bins including all 256 energy channels. We performed fast
Fourier transforms (FFTs) of 128 s data segments, fixing the frequency
resolution and the lowest available frequency to $1/128$ Hz;
the highest available Fourier frequency (Nyquist frequency) was 4096
Hz. No background subtraction or dead-time correction was made prior
to the FFTs.  The Poisson level was subtracted from the resulting
power spectra.  Following \citet{kle04} we first estimated the Poisson
noise using the \citet{zha95} formula and then (after inspecting the
high frequency range and finding no unexpected features) shifted it to
match the level between $\sim3000$--$4000$ Hz, where no intrinsic power
should be present, but only counting statistics noise. Then we
normalized the power spectra using the rms normalization
\citep{van95}.
 In this normalization, the integral over the power spectrum is
equal to the fractional rms amplitude squared.
The power density units are $\rm\,(rms/mean)^{2}Hz^{-1}$ and  the
fractional rms amplitude in one specific band is:
\begin{equation}
\rm\,rms=\it\left[\int_{\nu_{1}}^{\nu_{2}} P(\nu)\, d\nu\right]^{1/2}
\end{equation}

 The errors on the fractional rms are calculated by using the
dispersion of points in the data.  We consider a measurement as a
non-detection in a specific band, when the ratio between the
fractional rms and its standard deviation is smaller than 3. In this
case we quote upper limits at the 98\% confidence level.

The periodic 401 Hz pulsations were analyzed by constructing 
pulse profiles folding chunks of lightcurve
in profiles of $N=32$ bins, with the ephemeris of J1808 provided by
\citet{har09}. In this folding process we used the TEMPO
pulsar timing program (v 11.005) to generate a series of polynomial
expansions of the ephemeris that predict the barycentered phase of
each photon detected.  The length of each data chunk was chosen
according to the length of each {{\it RXTE}} observation (Obs-Id).

We then split each pulse profile into a fundamental (at $\nu$) and a
first overtone (at $2\nu$) using standard $\chi^{2}$ fits.  
For a detailed description of the method used for
measuring the pulse time of arrivals (TOAs) we refer to \citet{pat09}.
A set of pulse phase residuals was then obtained by subtracting a
Keplerian circular orbit and constant pulse frequency model, using the 
ephemeris of \citet{har09}.

\subsection{{{\it Swift}-XRT} observations}
During the 2005 and 2008 outbursts, the {{\it Swift} X-ray Telescope}
({{\it Swift}-XRT}) observed the re-flaring state of J1808.  We refer
to \citet{cam08} for the study of the {{\it Swift}-XRT} 2005
observations.  Here we consider 10 pointed observations (see Table 1)
covering a total of $\sim 15$ ks that were taken during the 2008
outburst (MJD$\,54757$--$54778$) and were reduced by using the XRT
pipeline (v. 0.12.0).  Each observation lasted between 1 and 3 ks.
The data were collected in photon counting (PC) mode, except for
observation 00030034041 which was taken in windowed timing (WT) mode.

We extracted source and background events for each observation using
circular regions with radii of 20 arcseconds, and extracting photons
with energies between 2--10 keV and 0.5--10 keV.

%########### TABLE ###########################
\begin{deluxetable}{ccccc}
%\centering
\tabletypesize{\footnotesize}
\tablecolumns{5}
\tablewidth{0pt}
\tablecaption{\textit{RXTE} and \textit{Swift}-XTE observations analyzed for each outburst}
\tablehead{
  \colhead{Outburst} &
  \colhead{Start} &
  \colhead{End} &
  \colhead{Time} &
  \colhead{Program IDs}\\
  \colhead{(year)}&
   \colhead{(MJD)} &
   \colhead{(MJD)} &
   \colhead{(ks)} & 
   \colhead{}
}
%\scriptsize
%\begin{tabular}{lrrrl}
%\hline
%\hline
\startdata
\RXTE & & & &\\
1998 & 50914.8  & 50939.6  & 160 & {\tt 30411}\\
2000 & 51564.1  & 51604.6  & 130 & {\tt 40035}\\
2002 & 52562.1  & 52604.8  & 664 & {\tt 70080}\\
2005 & 53522.7  & 53587.9  & 267 & {\tt 91056},{\tt 91048}\\
2008 & 54731.9  & 54776.1  & 236 & {\tt 93027}\\\hline
&&&&\\
SWIFT & & & &\\
%Outburst & Start & End & Time & Observation IDs \\
% (year) & (MJD) & (MJD) & (ks) &\\
2008 & 54756.6 & 54778.3   & 15 & {\tt 0003003435}-{\tt 00030034044}\\
\enddata
\label{tab:rxteobs}
\end{deluxetable}
%########### END TABLE ####################### 

\section{Results}\label{1Hz:results}

\subsection{The X-ray lightcurves and the re-flaring state}
\label{re-flarings}

In the 2008 outburst re-flaring state, the lightcurve reached very
faint luminosities, with a large portion of the re-flares below the
sensitivity limit of {{\it RXTE}}, but above the detection threshold
of {{\it Swift}-XRT}.  The flux reached a minimum level of $2.0\times
10^{-13}\rm\,erg\,s^{-1}\,cm^{-2}$ in the 0.5--10 keV band, which
corresponds to a luminosity of $\sim 3\times 10^{32} \rm\,erg\,s^{-1}$
at 3.5 kpc.  This luminosity is of the same order of magnitude as that
observed by \citet{wij03} and \citet{cam08} during the 2000 and 2005
re-flares.

\begin{figure}[!th]
  \begin{center}
    \rotatebox{0}{\includegraphics[width=1.0\columnwidth,clip]{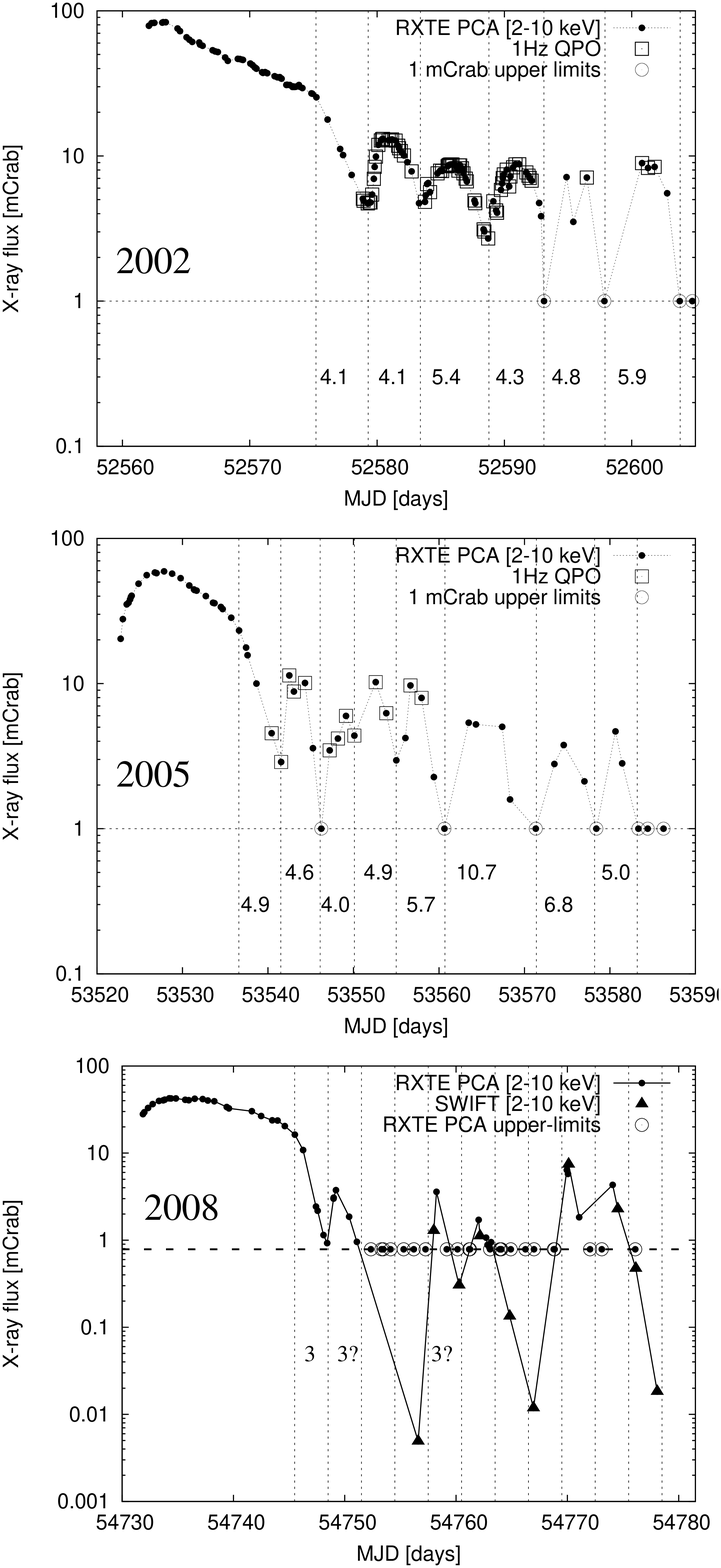}}
  \end{center}
  \caption{Outburst lightcurves in the 2-10 keV energy band.  The
count rate was normalized to the Crab intensity measurements nearest
in time \citep{kuu94} and in the same PCA gain epoch (e.g.,
\citealt{str03}).  Each black circle and triangle corresponds to one
{{\it RXTE}} and{{\it Swift}-XRT} observation, respectively (see Table
1 for a complete list of the observations used). The open squares
indicate observations when the 1 Hz QPO was detected, while the open
circles indicate the {{\it RXTE}} observations with an X-ray flux
below the detection threshold of $\sim 1$ mCrab.  The vertical dashed
lines divide the lightcurve into intervals beginning with the fast
decay, and separating the re-flares. The duration of each re-flare is
indicated in days and is similar to that of the fast decay (also
indicated).  In the 2008 outburst only few observations were above the
detection threshold and all the vertical lines refer to the fast decay
time scale.
    \label{tics-02-08}}
\end{figure}

A property of all the re-flares is the periodicity on time scales of a
few days, creating the ``bumps'' observed in the lightcurve (see
Fig.~\ref{tics-02-08}).  The fast decay and the re-flares in the 2002
and 2005 show a similar duration. The 2008 outburst also shows
comparable timescales, although with much higher uncertainty. 
As noted in \S~\ref{1Hz:intro},
in '98 the observations stopped too early and the '00 ones started too
late to allow a similar comparison.

The precise determination of the re-flare periods suffers of biases
due to occasional poor sampling and to the sensitivity
limit of {{\it RXTE}}. Therefore, although
the data certainly allow this, there is no significant evidence that
differences in fast decay time scales between the 3 outbursts have an
effect on the subsequent re-flare periodicity time scales.

In the 2008 outburst it is hard to calculate the re-flare time scale
because the flux is below the sensitivity limit of {{\it RXTE}} in the
majority of the observations. The {{\it Swift}} sampling was
approximately 1 observation every 2 days, too long compared to the
fast decay time scale of 3 days to unambiguously exclude shorter
time scales. However, the fast decay and the re-flare time scales are
again compatible with being the same.

In 2002 and 2005 the re-flares' peak luminosity tends to decrease with
time (see Fig.~\ref{tics-02-08}).  This is not observed in the 2000
(see \citealt{wij01})
and 2008 re-flares, which show an erratic change of peak luminosities. 
In the re-flaring state, the
luminosity can change by $\sim 3$ orders of magnitude on time scales of
$\simless 1.5$ days (see Fig.~\ref{tics-02-08} at MJD$\sim54756$ and
\citealt{wij03} for a similar observation in the 2000 outburst).

\subsection{The fast decays and the pulse phase drifts}
\label{fastdecayjump}

Close to the end of the 2002 outburst, the pulse phase of the fundamental was
observed to drift by $0.2$ cycles in just a few days, in coincidence
with the beginning and the end of the fast decay
\citep{bur06}.  \citet{har08} showed that a similar drift was present
in 2005. Here we show that both phase drifts start and end in
coincidence with (or very close to) the beginning and the end of the
fast decay.

\begin{figure*}[!th]
  \begin{center}
    \rotatebox{0}{\includegraphics[width=2.0\columnwidth]{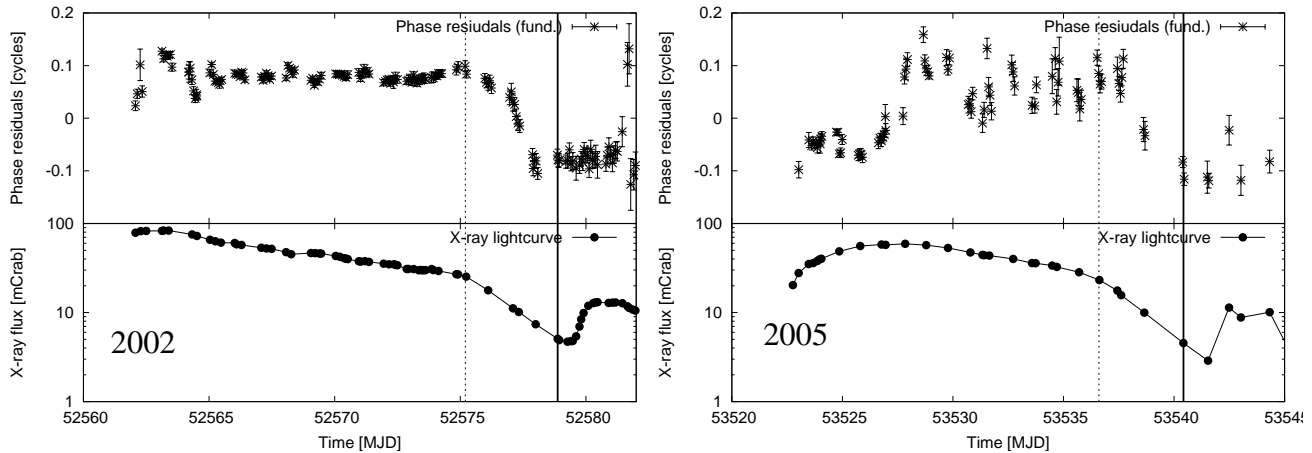}}
  \end{center}
  \caption{Pulse phase residuals (upper panels)
 calculated for the fundamental
frequency alone. The bottom panels show the X-ray lightcurve (in
logarithmic scale).  The dashed lines mark the beginning of
the fast decay in the lightcurves. At this point the pulse phases
start to drift for $\sim0.2$ spin cycles. The solid black lines mark the
time of the first appearance of the 1 Hz QPO.  Note that in both
outbursts it appears at the end of the phase drift close to (but
before) the minimum flux level in the X-ray lightcurve.
    \label{decay}}
\end{figure*}

In Fig.~\ref{decay} we plot the pulse timing residuals of the
 fundamental and the 2-10 keV X-ray flux in mCrab.  In the 2002
 outburst the pulse phase drift starts exactly when the fast decay
 begins. 
In the 2005 outburst we see the phase drift beginning close
 to the beginning of the decay, although given the large scatter in
 the pulse phases it is difficult to define the exact moment when the
 phases begin to drift.  Both the 2002 and 2005 pulse phase drifts end
 in coincidence with the end of the fast decay.  Both the 2002 and
 2005 pulse phases drift by $\sim0.2$ cycles.

\begin{figure*}[!th]
  \begin{center}
    \rotatebox{-90}{\includegraphics[width=0.5\textwidth]{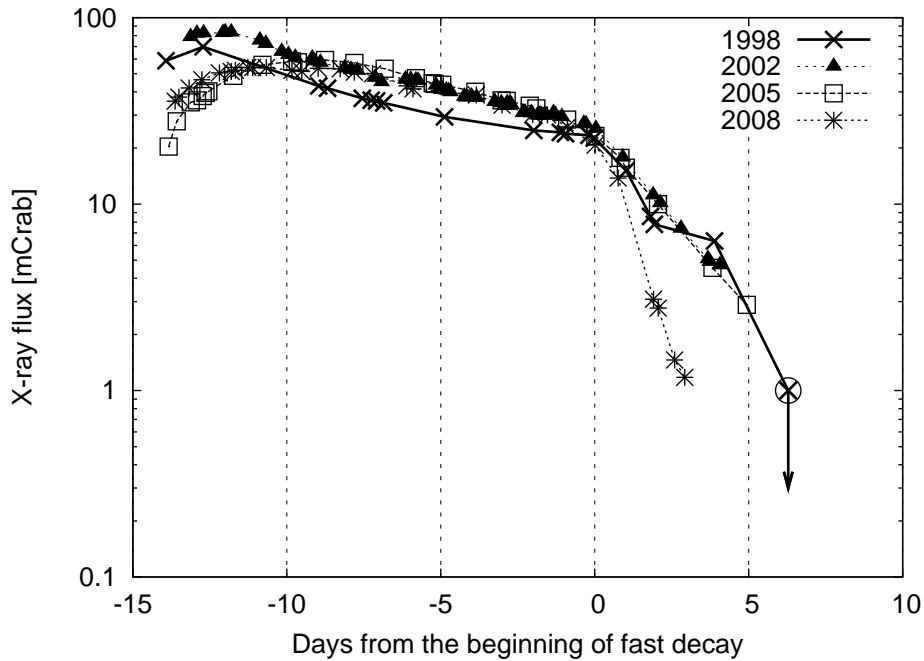}}
  \end{center}
  \caption{X-ray lightcurve of four outbursts, plotted up to the
end of the fast decay stage.  The curves are aligned to the beginning
of the fast decay.  The 2008 outburst has the shortest decay time.
The 2008 slow decay, peak and fast rise stages are very similar to
those of the 2005 outburst. The end of the 1998 outburst fast decay is
limited by a non detection (open circle); it is the outburst that shows
the dimmest luminosities before the beginning of the re-flares.
    \label{all-lc}}
\end{figure*}

The slope of the fast decay is consistent with being the same in
both outbursts (see also Fig.~\ref{all-lc}).  The pulse phases in both outbursts
during this time drift at the same speed of $\sim 0.07\rm\,cycle/day$.
The pulse phase behavior in the 2002 and 2005 fast decay is
therefore consistent with being identical.

In the 2000 outburst a similar test cannot be performed since {{\it RXTE}}
missed the decaying portion of the outburst, and the observations
covered just the re-flaring state. In the 1998 and 2008 outbursts no clear
phase drift is observed. We refer to \citet{har08, har09} for a detailed
discussion of the coherent timing analysis of those two outbursts.

\subsection{QPO parameters and flux}\label{phenomenology}
\label{qporelations}

We will not in this paper provide a complete description of all the
aperiodic variability observed in the 5 outbursts. For a description
of the aperiodic timing features observed in AMXPs, we refer to
\citet{str05} and to the review of \citet{wij06}.  Here we focus our
attention on the 1 Hz modulation and give a brief description of the
power spectra observed from the fast decay stage on, in order to
provide context. A description of our quantitative analysis of the 1Hz
modulation follows in the next sections.

\begin{figure*}[!th]
  \begin{center}
    \rotatebox{-90}{\includegraphics[width=0.5\textwidth]{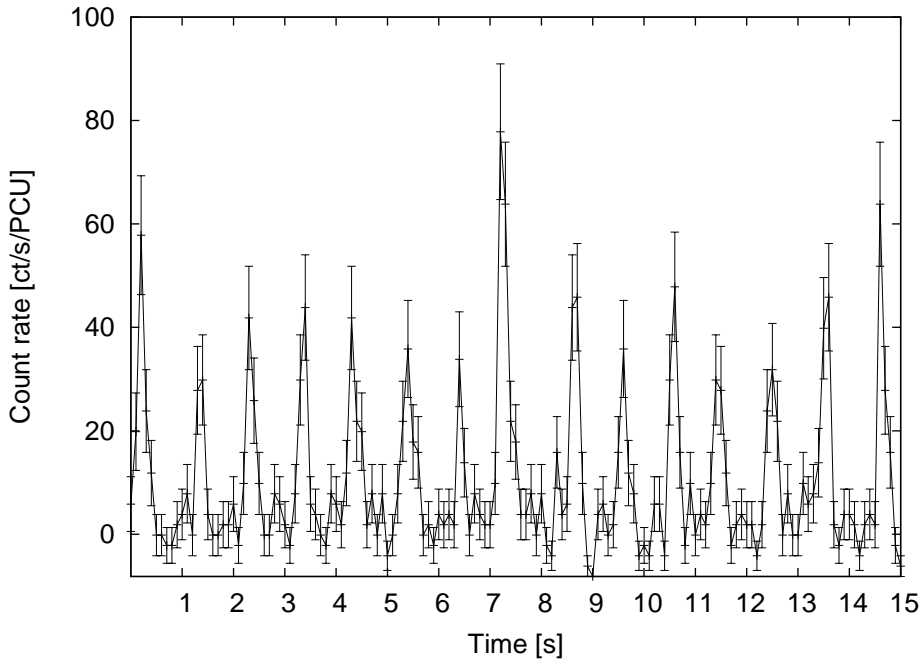}}
  \end{center}
  \caption{1 Hz modulation of the X-ray lightcurve (background
subtracted), as observed in a re-flare of the 2005 outburst. The figure
shows a 15 s chunk of lightcurve with a time resolution of 0.1
seconds. In this observation the 1 Hz modulation has the highest
measured fractional rms amplitude (125\% rms, ObsId:91418-01-02-05).
    \label{modulation}}
\end{figure*}

\begin{figure}[!th]
  \begin{center}
    \rotatebox{0}{\includegraphics[width=1.0\columnwidth,clip]{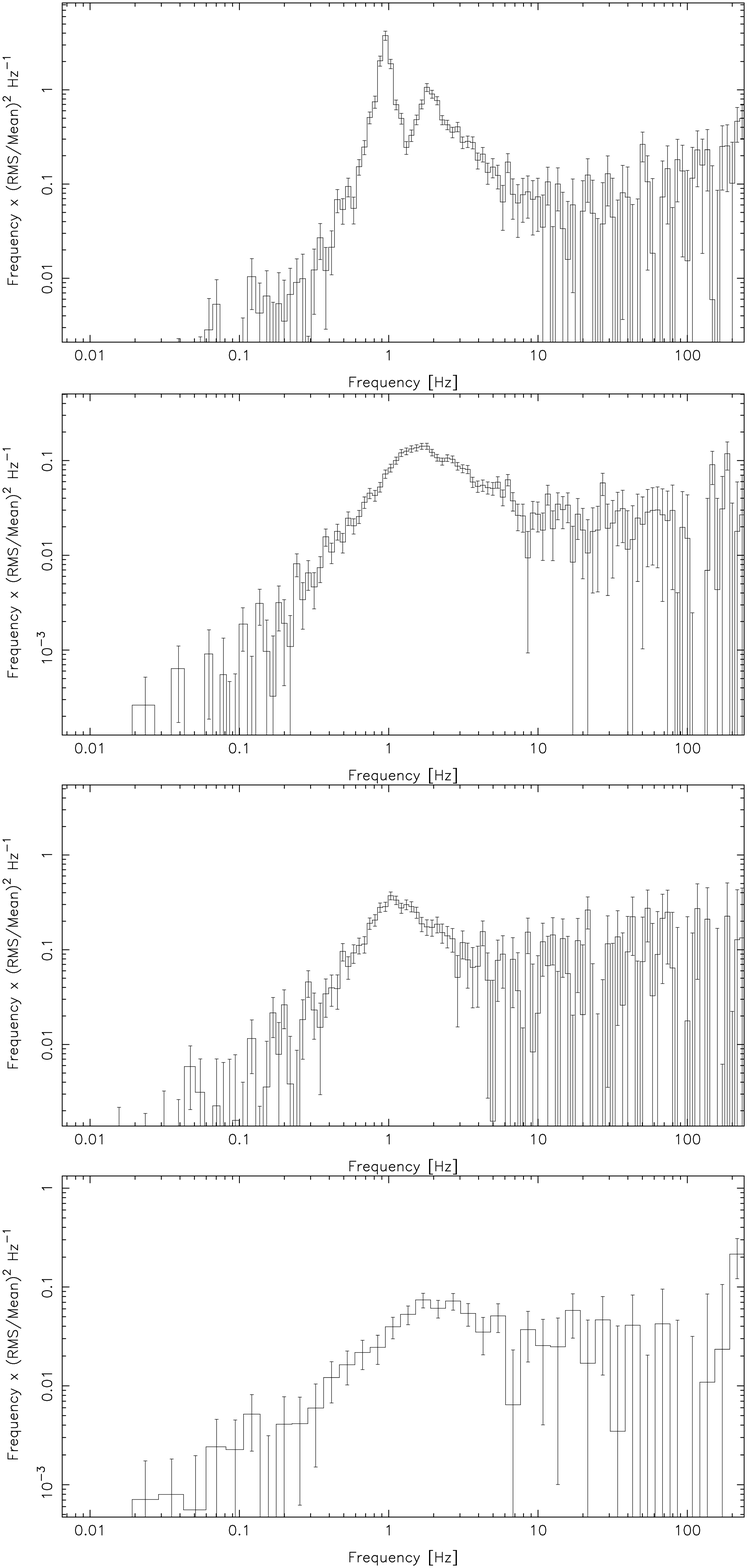}}
  \end{center}
  \caption{Power spectra (power$\times$frequency) of the re-flares of the 2005
outburst.  The four plots show different manifestations of the 1 Hz
modulation in the power spectrum.  The upper panel
(ObsId:91418-01-02-05, MJD$\sim53550.1$) shows a strong coherent 1 Hz
QPO and its overtone clearly separated.  They blend together in the
second panel (ObsId: 91418-01-01-00, MJD$\sim53542.5$) where some power
is also observed at frequencies higher than 10 Hz. The third panel
(ObsId: 91056-01-04-01, MJD$\sim 53541.5$) shows a QPO with a steep
cutoff at frequencies larger than $\sim 10$ Hz.  The bottom panel
(ObsId: 91418-01-03-04, MJD$\sim53556.7$) shows still power, but as an
incoherent feature without a clear peak.
    \label{QPO}}
\end{figure}

We examined the entire data set of all 5 outbursts for evidence of
the 1 Hz modulation. Close to the end of the 2002 and 2005 pulse
phase drifts, when the fast decay state is almost over, the X-ray
lightcurve clearly shows this strong modulation with a repetition frequency of
$\sim1$ Hz (Fig.~\ref{modulation}). It shows up as a
strong quasi periodic oscillation (QPO) peak around 1 Hz in the power
spectrum (Fig.~\ref{QPO}, see also \citealt{wij04} for a similar plot
for the 2002 outburst).
Later during the re-flares of these outbursts the modulation occasionally
recurs, as discussed in more detail in \S~\ref{1Hz:appearence}.
The 1 Hz modulation also appears and disappears sporadically during
the re-flares of the 2000 but was not detected in the 1998 and 2008
outbursts (cf. \S~\ref{1Hz:intro}).

We calculated the fractional rms amplitude\footnote{Fractional rms
amplitude is standard deviation divided by mean flux. Values of rms
larger than 100\% indicate that the 1 Hz flares have a large amplitude
and a short duty cycle For a light curve composed exclusively of
square flares with duty cycle $f$,
$\rm\,rms=\it\,\sqrt{\frac{1-f}{f}}$, arbitrarily large for
$f\rightarrow 0$} of the 1 Hz QPO by integrating the power in the 9.95
Hz wide band $0.05-10\rm\, Hz$.  This frequency band was chosen
because it is here that the power is observed in the large majority of
the observations.  Some power above 10 Hz is observed in some cases as
an extended tail of the 1 Hz QPO (Fig.~\ref{QPO}), but selecting an
upper frequency of $20\rm\, Hz$ does not change the results
significantly.  Clearly, using this method we sum up all the power
from the different components in the power spectrum: power of the 1 Hz
QPO, its harmonics, and the broad band noise. So everything in the
range 0.05--10 Hz is included, and therefore this
method is not optimal to measure the power generated by one single
component.  However, the rms calculated in this way is empirical and
independent from any model used to fit the data.  The fractional rms
amplitude in the 0.05--10 Hz band was in the range 10$-$125\% in all
observations where the modulation was detected.

The QPO is usually quite broad, and its shape cannot be satisfactorily
fitted by single or multiple Lorentzians, as it has a sharp fall-off
in power at lower frequencies. In some observations a second harmonic
peak is visible (Fig.~\ref{QPO}).
We model these features with Gaussians of the form:
\begin{equation}
P(\nu_{i}) = A\cdot \rm\,exp\it\left[(\nu_{i} - B)^{2}/C^{2}\right]
\end{equation}
where $\nu_{i}$ are the Fourier frequencies, A is a normalization
factor, B is the centroid frequency ($\nu_{0}$ or $2\nu_{0}$) and C is
related to the FWHM
through:$\rm\,FWHM=2C\cdot\,\left[ln(2)\right]^{1/2}$.
The typical value for the FWHM of the two Gaussians is $\sim1$Hz for
all the three outbursts.  The Gaussian that we use to fit the 1 Hz
modulation usually has a quality factor $Q=\nu_{0}/\rm\,FWHM < 2$.
When $Q<1$ the power spectrum does not show a clear peak, but
sometimes it does show a break in the noise (see for example the
bottom panel in Fig.~\ref{QPO}).  Here we call the 1 Hz modulation
``QPO'' regardless of its quality factor. When $Q<1$ we did not try to
fit the power spectra with gaussians because the centroid frequency is
ill defined. When the 1 Hz QPO is fitted by 2 Gaussians $\nu_{0}$ is
defined as being the frequency of the Gaussian with the highest peak
power in the power spectrum. With this choice $\nu_{0}$ is consistent
with being always the fundamental frequency of the 1 Hz modulation
(varying between 0.8 and 1.6 Hz), as the harmonic peak at $2\nu_{0}$
is always lower in maximum power.

\begin{figure*}[!th]
  \begin{center}
    \rotatebox{0}{\includegraphics[width=1.0\textwidth]{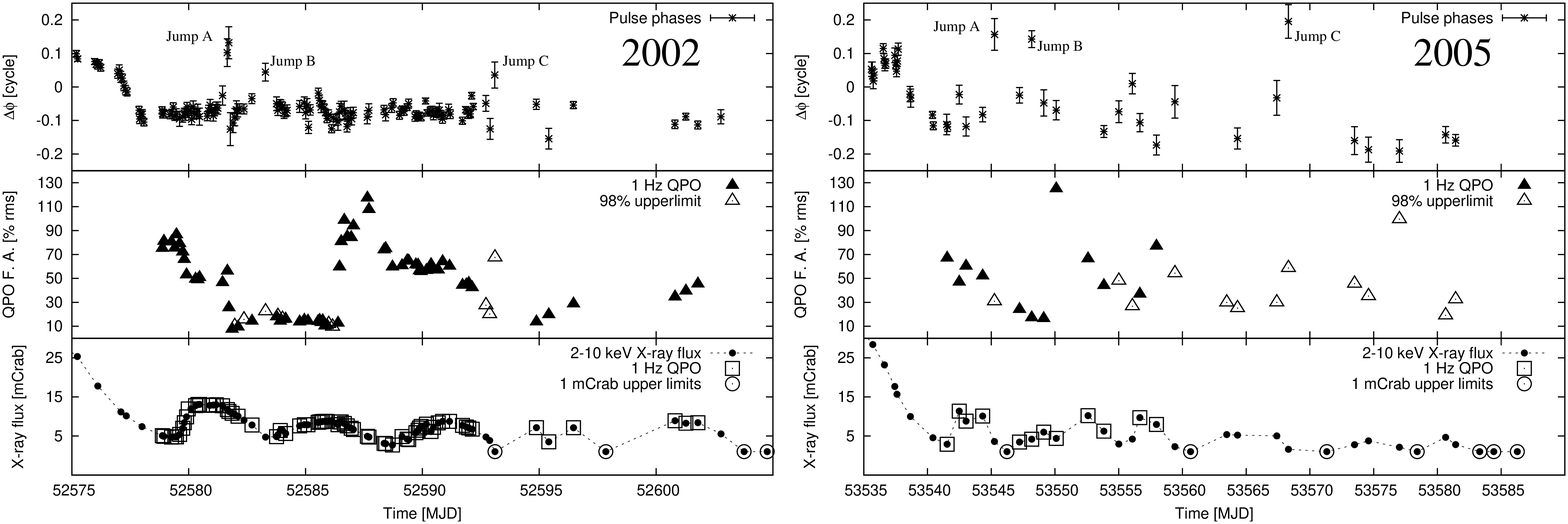}}
  \end{center}
  \caption{\emph{Upper panels}: pulse phases of the fundamental for
 the 2002 and 2005 re-flaring states.\\ \emph{Central panels}:
 Fractional rms amplitude of the 1 Hz QPO (black triangles) and
 non-detections at 98\% confidence level (open triangles).\\
 \emph{Bottom panels}: 2-10 keV X-ray lightcurves. The observations where the 1 Hz QPO is
 significantly detected are marked with open squares, while measurements 
where the X-ray flux was below the sensitivity limit are marked with open
 circles. The pulse phase shows jumps of 0.1-0.2 cycles on a
 time scale of hours or less, when the 1 Hz QPO disappears or when its rms
 amplitude reaches a low level.
    \label{jumps}}
\end{figure*}

We calculated the fractional rms amplitude and centroid frequency of
the 1 Hz QPO and the X-ray flux for the 2000, 2002 and 2005 outbursts.
The fractional rms amplitude as observed in 2002 and 2005 is shown in
Fig.~\ref{jumps} (middle panel). The fractional rms amplitude shows abrupt
changes over time and does not follow a clear correlation with flux.
In the 2000 re-flaring state, the 1 Hz modulation was detected in
 13 observations out of  46, close to the peaks of the re-flares.

\begin{figure*}
  \begin{center}
    \rotatebox{0}{\includegraphics[width=2.0\columnwidth]{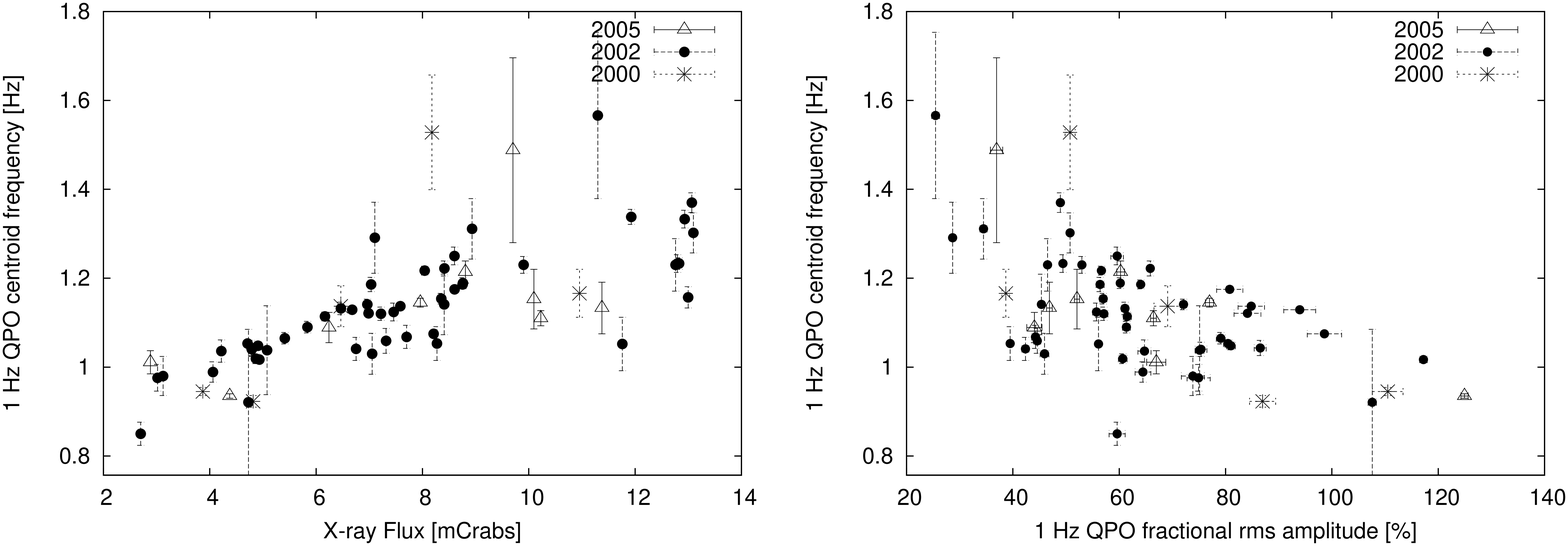}}
  \end{center}
  \caption{ 1 Hz QPO centroid frequency vs. X-ray flux (left), 1 Hz
 QPO centroid frequency vs. 1Hz QPO fractional rms amplitude (right
 panel) for the 2000 (asterisks), 2002 (circles) and 2005 (open
 triangles) outbursts.  The QPO centroid frequency is clearly
 correlated with the X-ray flux and increases with decreasing
 fractional rms. The two relations are similar 
 in all three outbursts.
    \label{rms-flux-freq}}
\end{figure*}

We plotted together all the points for the three
outbursts in Fig.~\ref{rms-flux-freq}.
There is a clear increase of the frequency with flux and
and an anti-correlation between frequency and QPO fractional rms
amplitude. Both relations have considerable scatter.  
We performed a rank correlation test on the 1 Hz QPO frequency vs. 
fractional rms anticorrelation and we found a Spearman coefficient
of $\rho=-0.38$ with a probability of 0.1\% of the null hypothesis
(no correlation in the data) being true. 
A similar test for the frequency vs. flux correlation gives $\rho=0.75$ 
with a probability of less than 0.01\% for the null hypothesis.

We explored the fractional rms dependence of the 1 Hz QPO on the X-ray
flux using a similar plot, and we found no clear dependence. 


We then calculated the upper limits (quoted at the 98\% confidence
level) for all the observations after the end of the fast decay where
the QPO was not detected.  The upper limits were calculated per
observation.  
In the 1998 data two complications occur: the observations ended
immediately after the beginning of the re-flares, and the fast decay
reached very faint fluxes ($< 1$ mCrab) at its minimum.  Only in one
observation J1808 was detected; then the upper limit was 55$\%$ rms. A 1
Hz modulation therefore cannot be completely excluded for this
outburst.  In 2000, 2002, 2005 and 2008 the most constraining upper
limits are 15$\%$, 9$\%$, 19\% and 19\% rms, respectively.

Across the 2000, 2002 and 2005 outbursts, the 1 Hz QPO was observed in a rather
narrow range of luminosities during the re-flares ($2$--$15$ mCrab,
2--10 keV), in contrast with the larger range of luminosities covered
by the outbursts ($\sim 1$--$80$ mCrab). The {{\it Swift}-XRT} and {{\it
XMM}-Newton} observations had insufficient time resolution or an
insufficient number of counts to probe the presence of the 1 Hz QPO
below 1 mCrab.

\begin{figure}[!th]
  \begin{center}
    \rotatebox{-90}{\includegraphics[width=0.7\columnwidth]{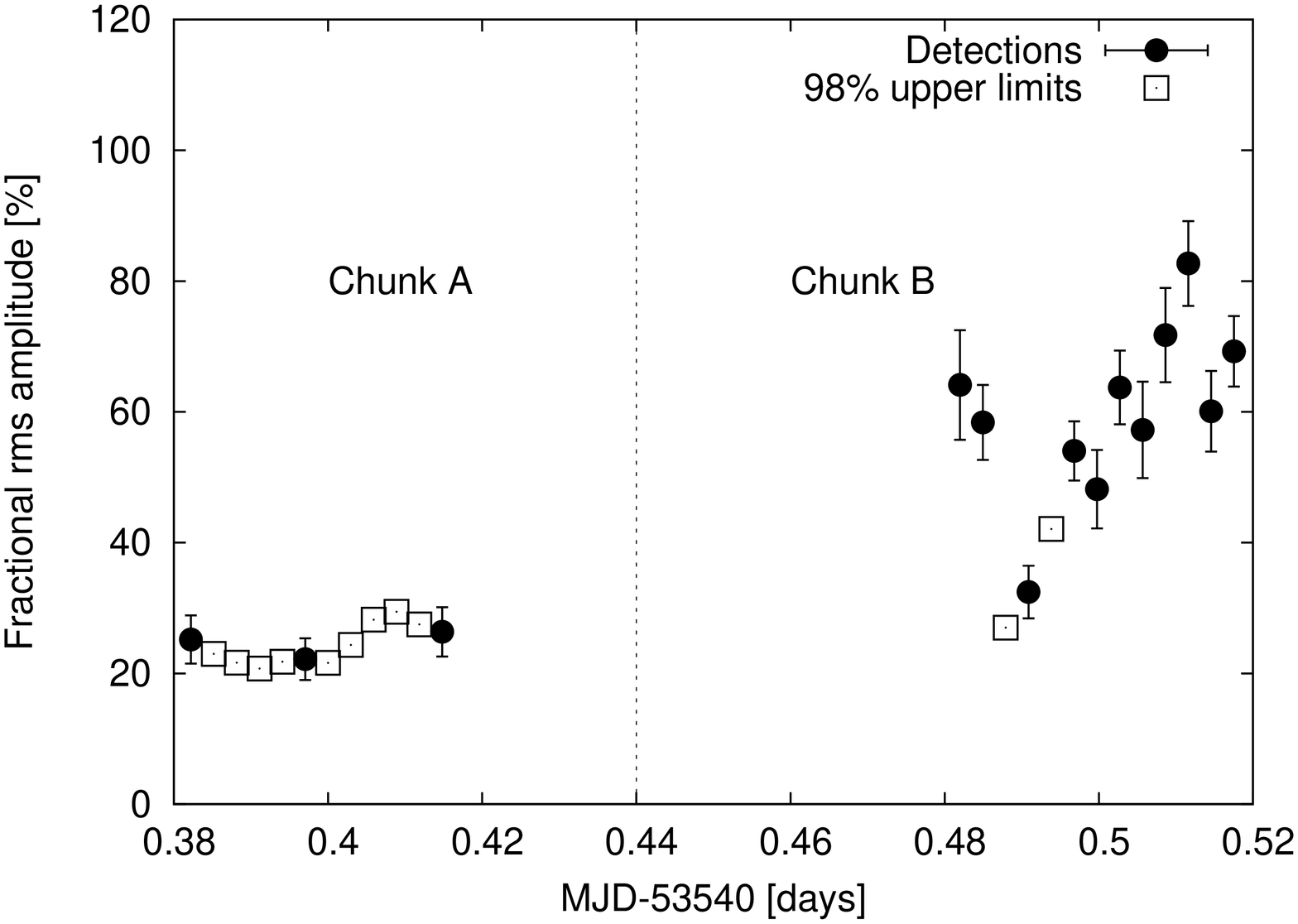}}
  \end{center}
  \caption{Temporal evolution of the 0.05-10 Hz fractional rms
amplitude in observation 91056-01-04-03, showing the 1 Hz QPO
appearance during the 2005 outburst.  Each point corresponds to a 256
s long observation.  Chunks A and B correspond to two different {{\it
RXTE}} orbits. The black circles are $>3\sigma$ detections, while the
open squares are 98\% confidence upper limits.  The rms amplitude of
chunk A is always below $30\%$ rms, with 3 significant detections
where the 1 Hz QPO is a broad incoherent feature in the 0.05-10 Hz
band. In chunk B, the 1 Hz QPO fractional amplitude can reach values
of up to $\sim 85\%$ rms and it fluctuates on very short time scales
($\sim256$ s).
    \label{appears}}
\end{figure}

\subsection{The appearance of the 1 Hz QPO}\label{1Hz:appearence}

In the 2002 outburst, the 1Hz QPO appears immediately after the pulse
phase drift is complete, but with the flux still decreasing in the
fast decay stage.  In 2005 the 1 Hz QPO appears in a similar position
(see black vertical lines in Fig.~\ref{decay}).  In both the 2002 and
2005 outbursts, data gaps prevent the observation of the exact moment
when the 1 Hz QPO appears.  Taking account of the gaps, the flux level
of the first 1 Hz QPO appearance is consistent between the two
outbursts.

To track the appearance of the 1 Hz QPO on time scales as short as a
few hundred seconds, we investigated the observation available in
which the 1 Hz QPO first appears in the 2005 outburst.  The
observation (Obs-Id 91056-01-04-03) shows two chunks of data separated
by a gap of 5000 s (chunk A with a length of 3200 s and chunk B with
length $3500$s) that we analyzed separately.

In chunk A the power spectrum has already changed with respect to the
earlier observations: the power at higher frequencies has disappeared,
and there is only power in the range 0.05--10 Hz. Probably we are
witnessing the onset of the 1 Hz modulation.  The average power
spectrum shows a fractional amplitude of $(21\pm 1)\%$ rms in the
0.05--10 Hz band.
% (Fig~\ref{noQPO}).  
In chunk B we find a 1 Hz QPO
clearly present with a fractional amplitude of $(68\pm4)\%$ rms.

We then calculated power spectra from 256 s data segments and
 determined the fractional rms amplitude in the 0.05-10 Hz band.  (see
 Fig.~\ref{appears}). Significant power is detected in three segments
 during the observations of chunk A, meaning that some low level
 activity is already present in the same frequency range where the QPO
 will later appear. In chunk B power is nearly always detected; the
 onset of the 1 Hz QPO must have occurred within the 5000 s gap. At
 MJD~53540.48 the amplitude shows an abrupt decrease from $\sim 60\%$
 rms down to $<30\%$ (98\% confidence level) on a time scale of $256$
 s.

We also calculated the power in a number of frequency bands in the range
1/128 to 256 Hz for all power spectra of chunk A.
The power is always consistent with zero at the three sigma level in
all the frequency bands except for the 0.05-10 Hz band in the three
segments mentioned above.

\subsection{Energy dependence of the 1 Hz QPO} 

\begin{figure}[!th]
  \begin{center}
    \rotatebox{-90}{\includegraphics[width=0.7\columnwidth]{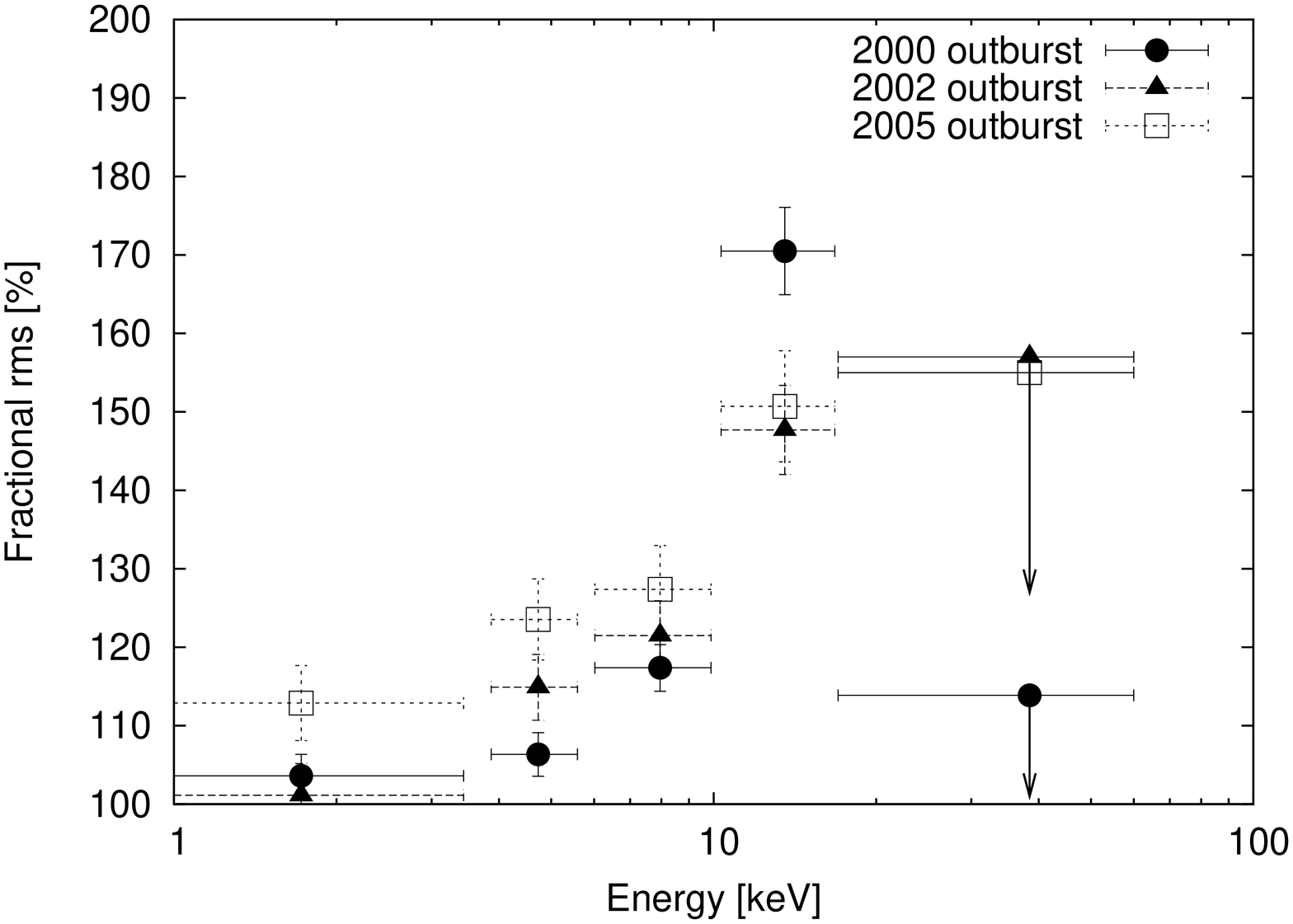}}
  \end{center}
  \caption{Energy dependence of the 1 Hz QPO fractional rms amplitude.
The rms increases with energy for all the three outbursts. The plot
shows rms amplitudes up to 17 keV, where a significant amount of
counts is detected. In the range 17-60 keV only upper limits (98\% 
confidence level) can be
calculated for the three outbursts.
    \label{energy}}
\end{figure}

In Fig.~\ref{energy} we show the energy dependence of the 1 Hz QPO.
The points in the figure refer to the observation (one per outburst)
for which the fractional rms of the 1 Hz QPO in the 2-60 keV energy
band was the highest ($110.5\%$ rms in 2000, ObsId 40035-01-04-01,
total exposure 7 ks; $117\%$ rms in 2002, ObsId 70080-03-15-00, total
exposure 2 ks; $125\%$ rms in 2005, ObsId 91418-01-02-05, total
exposure 1.2 ks). In these observations the characteristic
frequencies of the 1 Hz QPO and its overtone form clear peaks in the
power spectrum and $\rm\,Q>2$.

The energy spectrum is hard, rising by a factor of 1.5-1.7 between 2
and 17 keV.  Upper limits (98 $\%$ confidence level) in the 17-60 keV
band were: $114\%$ rms for 2000, $157\%$ rms for 2002 and
$155\%$ rms for 2005.

\subsection{Jumps in 401 Hz pulse phases related to strength of 1 Hz QPO}
\label{phasejump}

During the first re-flare of the 2002 outburst, the 1 Hz QPO becomes
very broad and then disappears.  When the QPO becomes very broad and
its fractional rms amplitude suddenly drops from $\sim 50\%$ down to
$10\%$, a jump of 0.2 cycles is observed in the pulse phases of the
fundamental (\textbf{jump A}, see Fig.~\ref{jumps}).  Two similar
pulse phase jumps of 0.1 cycles are observed
%$\sim 1.7$ days later
(\textbf{jump B} and \textbf{C}) when the QPO is not detected, with
rms amplitude upper limits of $\sim30$ and $\sim10$\% rms (98\%
confidence level).  We call these phase changes ``jumps'' as
opposed to the ``drifts'' observed during the fast decay stage
(\S~\ref{fastdecayjump}), since the $0.1-0.2$ cycle
jumps are sudden, occuring on time scales of a few
hours or less, while the 0.2 cycle drifts take $4-5$ days.

In the 2005 outburst we see something very similar, although the
re-flares have a sampling that is much worse: the observations are
rather sparse and are separated by at least 1 day.  
In both the 2002 and 2005 outbursts, the pulse phase in the jumps
is close to the pulse phase prior to the beginning of the fast
decay stage (see \S~\ref{fastdecayjump}).

In the 2000 outburst the observations are rather sparse, with gaps of 
several days between observations. However also in this case when 
the QPO rms amplitude is low or the QPO is undetectable, the pulse phases
are observed to jump by $\sim0.1$--$0.2$ cycles, consistent 
with the 2002 and 2005 pulse phase behavior.
 
This phenomenology relating the 1 Hz QPO rms amplitude and the pulse
 phases might suggest a link between the presence of the 1 Hz QPO and
 the accretion flow onto the neutron star surface.  However, due to
 the sparse sampling of the observations and the small number of jumps
 observed, this link remains to be confirmed in future observations.

\subsection{Dependence of 401 Hz pulsation on the 1 Hz phase}\label{peakvalley}

The 1 Hz modulation might affect the magnetic channeling
during the re-flaring state.
Possible differences in the pulse properties can be unveiled by
analyzing the pulsations coming from different phases of the 1 Hz
modulation. Since the modulation has a frequency that changes with
time, it is hard to unambiguously define its phase.

We therefore used a crude but simple approach, defining ``peaks'' and
``valleys'' of the 1 Hz modulation in the lightcurve relative to the
average flux.  We first calculated the average X-ray flux per
observation.  Then we split the X-ray lightcurve into an upper half and a
lower half containing peaks and valleys,
respectively.  We checked that this method was properly dividing the
lightcurve by a visual inspection.  We then folded all the peaks and
the valleys into two different 401 Hz pulse profiles at the spin
frequency of the neutron star (see \citealt{har08} for the 
ephemeris used to fold the data), and repeated the procedure for each
observation where the 1 Hz QPO was detected.

The pulse phases are always consistent with being the same for peaks
and valleys, within the pulse phase uncertainty of down to 0.01
cycles.  The absolute pulse amplitude is higher in the peaks than in
the valleys but by a smaller fraction than the mean flux, so that the
pulse fractional amplitude is \textit{always} higher in the valleys by
a factor varying between 1.2 and 6 (with a typical error on the
amplitude ratios of $\sim0.1$).
The pulse amplitudes of peaks and valleys follow a similar evolution
in time: when the pulse amplitude of peaks rises (or drops) in an
observation, the same is seen for the pulse amplitude of valleys.

The valleys fractional rms amplitudes span a range between a few \%
rms up to more than 10 \% rms, with values much larger than the
average pulsed fractions of the persistent pulsation observed during
the outbursts before the reflares (for a discussion of the pulse
fractional amplitudes during the 1998, 2000, 2002 and 2005 outbursts
see \citealt{har08} and for the 2008 outburst see \citealt{har09}).
The opposite is true for the pulsed fractions of peaks: they are on
average smaller than the observed fractional amplitudes during the
outbursts.

\section{Discussion}\label{1Hz:Discussion}

Our analysis of the five outbursts of SAX J1808.4-3658 has shown that
the fast decay, the re-flares and the 1 Hz modulation can be related
phenomena in the 2002 and 2005 outbursts (and possibly in the 2000 one) 
which might also affect the 401 Hz pulsations.  The drifts
in pulse phase and the re-flares observed on time scales of days might
be related with some change in the disk structure which is connected
to the fast decay, which takes place on a similar time scale.  The
fast decay presumably is either a thermal or a viscous one, and
additionally depends on the lengthscale on which significant changes
occur within the disk, longer length scales corresponding to longer
time scales. The 1 Hz QPO properties had previously only been very
briefly discussed in the literature, and left the mechanism
responsible to be identified. We therefore studied the 1 Hz modulation
properties in detail and related them to the behavior of the X-ray
flux and the X-ray pulsations. Any suitable model for the 1 Hz
modulation has to explain the following key properties that we
reported in \S~\ref{1Hz:results}:
\begin{itemize}
\item{In 2002 and 2005, it appears at the end of the fast decay, after
the pulse phases have drifted by 0.2 cycles (Fig~\ref{decay})}
\item{It appears at a flux level that is consistent in the 2002 and
2005 outbursts (\S~\ref{fastdecayjump})}
\item{It appears sporadically only in a narrow range of luminosities ($\simless 15$mCrab,
\S~\ref{qporelations})}.
\item{It is very coherent in some observations while in some others it
is a broad incoherent feature (Fig.~\ref{modulation} and~\ref{QPO})}
\item{In a few cases, its amplitude might be connected with the 401 Hz pulse phases
(\S~\ref{phasejump}, Fig.~\ref{jumps}).}
\item{Its fractional rms amplitude can be as high as 125\% and then
$10\%$ for the same X-ray luminosity (Fig.~\ref{jumps})}
\item{Its amplitude is energy dependent, rising with energy up to $\sim 17$keV
(Fig.~\ref{energy})}
\item{Its rms amplitude is very high, up to 170\% in the 10-16 keV
band (Fig.~\ref{energy})}
\item{Its centroid frequency is quite stable and slightly increases
with flux (Fig.~\ref{rms-flux-freq})}
\item{Its centroid frequency decreases with increasing fractional rms
amplitude (Fig.~\ref{rms-flux-freq})}
\end{itemize}

Flares that might be similar to the J1808 re-flares are seen in other
low mass X-ray binaries (LMXBs): the black hole candidate XTE
J1650-500~\citep{tom03}, and the neutron star transient SAX
J1750.8-2900~\citep{lin08}.  This behavior therefore may not be unique
to J1808, although this is the only AMXP in which such re-flares have
been observed. 

However, the 1 Hz modulation does appear to be nearly unique to
J1808. The most similar QPOs, in terms of frequency and high
fractional amplitude, are found in the systems that show Type II
bursts.  The Rapid Burster \citep{lew76, hof78, mar79} has QPOs in the
range 0.04--4 Hz in the persistent emission, and from 2--5 Hz in
during its Type II bursts \citep{taw82, lew87, ste88, dot90, lub91,
lub92, rut95}. Similar frequencies (0.04--0.4 Hz) have also been
reported in the aftermath of Type II bursts from the bursting pulsar
GRO J1744-28 \citep{kom97}.  This similarity may point to a
magnetosphere/disk mechanism (S~\ref{disk-interaction}).

Oscillations between 0.58 and 2.44 Hz were also reported in three
dipping sources, although with an energy independent rms amplitude of
below $12\%$ \citep{hom99, jon99, jon00}. Disk warping or shadowing
was suggested as the most likely mechanism for these QPOs (see \S~\ref{diskwarping}).

Alternatively, the 1 Hz QPO may be caused either by the accretion flow
(including possible occultation phenomena), or by some process that
occurs after matter arrives on the surface.  If the latter, it is very
hard to understand why it is only seen in J1808.  A brief
consideration of the two most plausible surface mechanisms, global
oscillations or marginally stable nuclear burning, does in any case
seem to rule them out.

There are various oscillatory modes of the neutron star surface layers
that could lead to periodic brightness variations.  The most likely
candidates would be an ocean g-mode (a vibration driven by thermal
buoyancy \citealt{mcd87, mcd88}), although to obtain a 1 Hz frequency
would necessitate a very high order harmonic\citep{bil96}. The dependence of
amplitude on energy could be explained by a mode model, but modes
cannot explain the extremely high fractional amplitudes of the 1 Hz
QPO \citep{pir06}.  In addition, if this was the right mechanism, then
it should also be observed in other stars since the triggering
conditions should not be unique to J1808.

The matter accreted onto the surface of the neutron star may burn
stably, unstably (generating X-ray bursts), or in a marginally stable
fashion (for a review see \citealt{bil98b}). The time scale for the
quasi-periodic variations associated with marginally stable burning is
set by the accretion time scale and the thermal time scale. For
hydrogen ignition, the time scale for marginally stable nuclear
burning, computed as in \citet{heg07} using the appropriate accretion
and thermal time scales, is clearly too slow ($\sim 10$ minutes). For
helium ignition, the time scales are shorter but still too slow
($\sim100$ s), thus ruling out the marginally stable nuclear burning
scenario. In the following sections we therefore focus on the 
accretion flow as the most likely mechanism.

\subsection{Accretion onto a magnetized neutron star}
\label{genacc}

In order to understand what might be causing the 1 Hz QPOs, some
general background on magnetically channeled accretion will be useful.
To sustain channeled accretion at the maximum accretion rate of a few
percent Eddington, the magnetic field for J1808 must be $\gtrsim
4\times10^{7}$ G \citep{har08}. The upper limit on the field,
determined from timing, assuming that the spin down comes from
magnetic dipole radiation from a rotation powered pulsar, is
$1.5\times 10^8$ G \citep{har08, har09}.

In a discussion of magnetized accretion, reference is often made to
the corotation radius $r_c$, the radius at which matter in a Keplerian
orbit would have the same angular velocity as the star.  

\begin{equation}
r_c \sim 17 \left[\frac{\nu_s}{1 {~\rm kHz}}\right]^{-2/3}
\left[\frac{M}{1.4 ~M_\odot}\right]^{1/3} {\rm km} 
\label{rc}
\end{equation}
where $\nu_s$ is the spin frequency of the neutron star and $M$ is its
mass.  For J1808, assuming a mass of 1.4$M_{\odot}$, $r_c \sim 31$ km.  
The other radius of relevance is the
magnetospheric radius $r_m$:  the radius at which the magnetic field becomes
dynamically important in controlling the inflow of matter.
For spherically symmetric accretion, one can estimate $r_m$ by setting
the magnetic pressure equal to the ram pressure of free fall
\citep{lam73}:

\begin{eqnarray}
r_m & \sim & 7.8 \left[\frac{B}{10^8 {~\rm G}}\right]^{4/7}
\left[\frac{R}{10{~\rm km}}\right]^{12/7}
\left[\frac{M}{1.4~M_\odot}\right]^{-1/7} {}\nonumber \\ & & \times
\left[\frac{\dot{M}}{\dot{M}_{\rm Edd}}\right]^{-2/7} {\rm km}
\label{rm}
\end{eqnarray}
where we have assumed a dipole field $B\sim\mu/r^3$, $\mu$ being the
magnetic moment.  $R$ is the stellar radius and $\dot{M}$ the
accretion rate.  In the case where accretion occurs from a disk this
expression will be slightly modified by the rotational energy
of the disk \citep{spr93}.  Rotation of the
central star can also affect the location of $r_m$ \citep{lov99}.  

This simple order of magnitude estimate yields 18 km for SAX J1808 at
the peak of the outburst (assuming $\dot{M}=5\%\dot{M}_{Edd}$ and $B=10^{8}\,$G). 
This increases as the accretion rate falls,
becoming comparable to $r_c$ once the accretion rate drops to 1\% of
the Eddington rate. The precise value of $r_{m}$ will depend on 
details of inner disk physics as discussed in \citet{psa99}.

When $r_m < r_c$ accretion should proceed without difficulty, with the
magnetic field channeling material out of the disk and onto the
magnetic poles \citep{pri72}.  Once $r_m > r_c$, however, the
situation becomes more complex, and the system is said to be in the
so-called ``propeller'' regime.  Initially it was thought that for
$r_m > r_c$ accretion would cease, with matter being expelled from the
system \citep{ill75}.
Further study by \citet{spr93} showed that $r_m$ actually has to
exceed $r_c$ by a reasonable margin for material to be
expelled. Steady accretion is in fact possible when $r_m > r_c$, even
though the neutron star should spin down. In this stage the inner
edge of the disk stabilizes near $r_c$ and the density at the inner
disk rises allowing angular momentum to be transferred outwards
\citep{spr93, rap04}.  This type of disk structure predicts spin-down
without requiring penetration of the disk by the magnetic field far
beyond $r_c$, as was proposed in early works on the topic
\citep{gho79b}.  In this paper we use the term ``propeller'' to define
the condition in which $r_m > r_c$. In the next Section we will
examine in turn the various mechanisms that might be responsible for
the observed 1 Hz QPO.

\subsection{Candidate Mechanisms: disk/magnetosphere instabilities}
\label{disk-interaction}

There are numerous ways of obtaining variability from the disk and its
interactions with the stellar magnetosphere.  Many have been explored,
however, as a means of explaining the kHz QPOs - and so have
frequencies that would be very far off 1 Hz.  For this reason we will
neglect many of the mechanisms for disk variability that have been
discussed in the literature (see \citealt{van06} for an extensive
review of these mechanisms) and focus on those that might have
frequencies in the right range.

\subsubsection{Disk obscuration}
  \label{diskwarping}

The dipping QPOs have fractional amplitudes of $\sim10$\%
\citep{hom99, jon99, jon00}.  In the dipping sources the energy
dependence of the QPO amplitude was flat, supporting the shadowing
hypothesis. In J1808, there is a clear energy dependence and the 1 Hz
QPO amplitude is much higher than that seen in the dippers.
Furthermore, there is no evidence for dipping in the X-ray lightcurve
of J1808. Several recent studies ( \citealt{cac09}, \citealt{pap09},
\citealt{ibr09}, \citealt{del08}) suggest an inclination for J1808 of
around $60^{\circ}$ which is too small to produce dipping.  The
mechanism proposed for this QPO was a partial obscuration of the
neutron star surface via a blob of plasma in the disk, orbiting at a
keplerian frequency of $\sim 1$ Hz.  The disk radius corresponding to
a keplerian frequency of 1 Hz is $r\sim 1700$ km. Considering the
observer at an inclination of $60^{\circ}$, the blob of plasma needs
to be as thick as 1000 km to allow partial obscuration of the neutron
star surface. The expected accretion disk thickness at that radius is
expected to be however orders of magnitude thinner than this value. By
using for example Eq. (9) in \citet{rap04}, the disk has an expected
scale height of $\sim 70$ km.  So, the 1 Hz QPO in J1808 is unlikely
to be explained with a dipping mechanism.

Another possibility is the occurrence of occulations
of the neutron star surface when the inner edge of the accretion disk
enters in the line of sight of the observer.
The 1 Hz QPO appears only at very low flux levels (in the range $\sim
0.001-0.01\%$ Eddington) which has two important consequences.  First,
by using Eq.~\ref{rm}, the inner edge of the disk is at approximately
56 km when the X-ray flux is at its minimum (0.001\% Eddington).  This
means that, by assuming an observer inclination of $60^{\circ}$, the
neutron star surface is obscured for a disk thickness of $\sim 60$ km.
This thickness is however much larger than the expected 
disk scale height which is only 0.04 km. 
Moreover, even considering the physical size of the inner disk 
as large as 60 km, the optical depth would drastically 
reduce to values below 1 after a few km in height 
from the disk middle plane.  
Although the inner part of the accretion disk could be optically thin, allowing
photoelectric absorption and thus explaining the hard energy
dependence of the 1 Hz QPO rms amplitude, it cannot explain the 
very high rms amplitude of the QPO, since the optically thin plasma 
cannot completely obscure the neutron star surface. 

\subsubsection{Interchange instabilities}
\label{interchange}

The role of interchange instabilities in admitting matter to the
magnetosphere is discussed in detail by \citet{aro76} and
\citet{els77}. The magnetic pressure prevents incoming matter from
crossing magnetic field lines, but if it is energetically favorable
(due for example to gravity) for material to be ``inside'' the field
lines rather than ``outside'', then interchange instabilities can act to
move the material inside.  This is often referred to as the
Rayleigh-Taylor instability, a term that refers to the instability
that occurs when a more dense fluid overlies a less dense fluid in a
gravitational field.  Where a plasma is supported against gravity by a
magnetic field, this is more correctly referred to as the
Kruskal-Schwarzschild instability \citep{kru54}.

The conditions necessary for the onset of Kruskal-Schwarzschild
instabilities in the situation where $r_m \ne r_c$ have since been
studied using MHD simulations \citep{rom08, kul08}.  For
misalignment angles $\theta \lesssim 30^\circ$ accretion can proceed
either stably via funnel flows, or unstably via interchange
instabilities.  In the unstable situation matter accretes via a number
of ``tongues'' that penetrate the magnetopause in the equatorial plane.
If a certain number of tongues dominate, quasi-periodic oscillations
can emerge in the light curves.  Funnel flows can co-exist with
accretion by tongues \citep{kul08}, although their presence should
reduce the amplitude of the persistent pulsations by reducing the
azimuthal asymmetry \citep{rom08}.  This might be consistent with the
continued presence of accretion-powered pulsations, with an
amplitude that depends on the phase of the 1 Hz QPO.

In the cases studied by \citet{rom08} interchange instabilities set in
above a critical accretion rate, making them unlikely as a cause for
the 1 Hz QPO (which appears only below a critical rate).  Interchange
instabilities cannot be completely ruled out, however.  In the
standard case studied by \citet{rom08} magnetic pressure and gravity
dominate the force equations.  When $r_m \sim r_c$, however,
magnetic pressure equals gravity.  At this point other terms start to
dominate the force equations and the character of the interchange
instability will change \citep{baa77, spr93}.  \citet{baa77} showed
that sporadic penetration of the magnetosphere is possible in this
regime.  However detailed numerical simulations of the type performed at
higher accretion rates have not been done, and the effect on funnel
flows (and hence the amplitude of the accretion-powered pulsations) is
not known.  Without further study, periodicity due to interchange
instabilities operating in the regime where $r_m \sim r_c$ cannot
be ruled out as a mechanism for the 1 Hz QPO.

\subsubsection{Magnetic reconnection instabilities}
\label{reconnection}

As matter moves within the disk (radially and azimuthally) it can drag
magnetic field lines along with it.  The sheared field lines can
temporarily impede accretion until reconnection establishes a normal
flow again. The resulting quasi-periodic accretion flow would lead to a
corresponding quasi-periodicity in the lightcurve, provided that the
accretion funnel and hot spot can respond on a 1 s time scale.

Magnetic reconnection is one of the most plausible mechanisms for Type
II bursts (cf. \S~\ref{1Hz:Discussion}).  Type II bursts are thought
to occur in systems where magnetic inhibition causes accreting matter
to build up in a reservoir outside the magnetosphere.  Once a
sufficient over-density of material has accumulated, instabilities
cause a catastrophic breach of the magnetospheric hammock, resulting
in sudden bursts of accretion \citep{lew76}. Dramatic changes in QPO
properties immediately before Type II bursts, with no detectable
change in the spectrum, have been interpreted as indicating that the
QPOs in the persistent emission are generated within the fuel
reservoir \citep{dot90}. To what extent this phenomenon is relevant to
our 1 Hz QPO is unclear. In the following discussion we explore
several mechanisms that can produce 1 Hz oscillations.

\citet{aly90} used reconnection to explain the QPOs observed in the
Rapid Burster, and discussed how the disk would be broken up in `blobs'
by magnetic reconnection instabilities. They predict a frequency a
few times the beat frequency between the rotation rate of the star
$\nu_s$ and the Keplerian frequency at the inner edge of the disk
$\nu_K$.  This implies that the frequency of the QPOs should pass
through zero at the point where $r_m \sim r_c$.

An argument in favor of this mechanism is that has been observed in
simulations (\citealt{goo97, goo99a, goo99b}, and \citealt{rom05}).
\citet{rom05} observed a strong quasi-periodicity associated with
outflows of matter when the system was in the propeller regime.
Unfortunately~\citet{rom05} do not model the emission mechanisms, so
whether outflows or discrete accretion would genuinely lead to high
amplitude QPOs in the X-ray emission is not clear. \citet{ust06},
using MHD simulations, showed that rapidly rotating accreting stars
have strong periodicity of this type linked to strong
outflows. Observationally, there is evidence for jet formation from
the radio detections in the decay of the 1998 outburst and the peak of
the 2002 outburst (\citealt{gae99}, \citealt{rup02}), although the
latter cannot be related with the propeller onset since it occurs at
the maximum fluxes observed for J1808.

The frequency, for this mechanism, should depend on accretion rate (as
observed, Fig.~\ref{rms-flux-freq}).  It should however occur at all
accretion rates: current models suggest no means of confining this
mechanism to a narrow range of accretion rates as we observe for the 1
Hz modulation (\S~\ref{qporelations}).  Another argument against this
mechanism is that this type of instability should occur independently
of field misalignment angle, so should be seen in all the AMXPs and
the non-pulsating LMXBs (assuming that they all have an external
magnetic field).

\subsubsection{Thermal/viscous/radiation-driven instabilities}
\label{thermalvisc}

The inner region of an $\alpha$ disk \citep{sha73} is unstable to
thermal and surface density perturbations \citep{pri73, lig74}. Both
types of instability can arise when radiation provides the major
contribution to the total pressure \citep{sha76}. 
\citet{fra02} derive the various time scales in operation:

\begin{equation}
\tau_\phi \sim \tau_z \sim \alpha \tau_{\rm th} \sim 10^{-4}
\left[\frac{M}{1~M_\odot}\right]^{-1/2}
\left[\frac{R}{10~{\rm km}}\right]^{3/2} s
\label{ttherm}
\end{equation}

\begin{eqnarray}
\tau_{\rm visc} & \sim & 3 \alpha^{-4/5}
\left[\frac{\dot{M}}{10^{16}~{\rm g/s}}\right]^{-3/10}
\left[\frac{M}{1~M_\odot}\right]^{1/4} \nonumber \\ && \times \left[\frac{R}{10~ {\rm
      km}}\right]^{5/4} s
\label{tvisc}
\end{eqnarray}
The time scales are defined as follows: $\tau_\phi$ is the dynamical
time scale in the disk; $\tau_z$ is the time scale on which deviations
from hydrostatic equilibrium in the z-direction get smoothed out;
$\tau_{\rm th}$ is the thermal time scale, that is the time scale for
re-adjustment to thermal equilibrium, if, say, the dissipation rate is
altered; and $\tau_{\rm visc}$ is the time scale on which matter
diffuses through the disk due to the effect of viscous torques. 
Note that 
the canonical value suggested by numerical simulations for $\alpha$ is
0.01. Thus the dynamical time scale is of the order $10^{-4}$ s, the
thermal time scale is 0.01 s and the viscous time scale is of the
order 100 s assuming that the appropriate length scale is comparable
to the inner disk radius $R\sim r_{c} \sim 10$ km. A study by
\citet{kin07} shows that for many LMXBs $\alpha \sim 0.1$, an order of
magnitude larger than the value suggested by numerical simulations of
disks. Therefore the observed time scales are expected to be shorter
than those calculated from numerical simulations. If the instability
were confined to a narrow inner annulus of the disk, then this might
be consistent with a 1 Hz frequency, since the viscous time scale
falls for shorter lengthscales. In addition, if the instability
triggers at the onset of the propeller regime when the inner edge of
the disk puffs up to sustain accretion, this might explain the rarity
of the phenomenon.

Instabilities which are known to operate at high accretion rates, such
as thermal-viscous instabilities in radiation dominated disks
\citep{taa84} and radiation-driven instabilities \citep{for89, mil95},
can be immediately ruled out since the 1 Hz QPO is observed at mass
accretion rates of $0.001-0.01\%$ Eddington.

An instability that might be relevant is the ionization
instability that is thought to put the system into outburst in the
first place (see \citealt{las01} for a detailed review). We know that
the 1 Hz QPO appears at the end of the outburst, where current
accretion disk models predict a transition from the hot to the cold
state.  The ionization
instability might enter a marginally stable oscillatory state 
when the disk
is on the verge of flipping between hot and cold regimes at
luminosities of $\simless 1\%$ Eddington. The question is then why it
would not occur always when the source is in the required luminosity
range, and why not also in other transient LMXBs. Fine tuning
by requiring this marginal state to coincide with for example 
the onset of the propeller regime and the associated changes in the 
disk structure as discussed above might resolve this. 
Therefore, the ionization instability, although unlikely, cannot be
ruled out.

\subsubsection{Spruit-Taam instability}
\label{spruittaam}

The Spruit-Taam instability involves radial perturbations at the
magnetospheric boundary (the inner edge of the accretion disk
according to \citealt{spr93}). It relies on the viscosity in the disk
to work, and does not require any shearing of the magnetic field.  As
discussed in Section \ref{genacc}, once $r_m \sim r_c$, accretion is
only possible if the density at the inner edge of the disk rises.
A small perturbation of the the disk radius away from $r_m = r_c$ will
be immediately damped and the disk radius will return to the 
``equilibrium'' position $r_m = r_c$ where inner disk edge and 
magnetosphere have the same angular velocity.

However, in the early propeller regime there exists a marginal state.
When the inner accretion disk is at equilibrium, a given $r_m$
corresponds to a specific density. If $r_m$ moves inward a little
bit, the boundary layer is 'over-dense' compared to the equilibrium for
that smaller $r_m$, and so quickly empties out. The rapid flow of
matter from the inner edge of the disk causes the rest of the
co-rotating transition zone to empty out too.  The rise in local
$\dot{m}$ pushes $r_m$ in still further, reinforcing the perturbation.
Eventually however the innermost layers are devoid of matter, since
the viscous time scale further out in the disk is too slow to have
replenished the inner regions. At this stage the boundary has to move
back outward. Matter then accumulates again until the cycle restarts.

The time scale on which this instability operates is related to the
viscous time scale just outside the corotating transition zone of the
disk, since this sets the time scale for replenishment of the
reservoir. However, this time scale is not related in a simple way to
the viscous time scale as it also depends on parameters like the
average accretion rate and the steepness of the transition between
disk and magnetosphere. 

According to eq.(\ref{tvisc}), the viscous time scale at $r_c$ for
J1808 is $\sim 100$ s, 2 orders of magnitude too long to explain the 1
Hz QPO. However this neglects the dependence of the time scale on the
additional parameters and assumes that the appropriate length scale is
comparable to the inner disk radius $r_{c}$. If a shorter length scale
were involved then a shorter viscous time scale would be possible
(\citealt{spr93}). To obtain a 1 second time scale, the length scale
of the region of activity would need to be $\sim1$ km, comparable to
the scale of the inner ``puffed up'' regions in the steady state disk
models calculated by \citet{rap04}.
The advantage of this mechanism is that the oscillatory state is
expected to be active only in a very narrow range of radii from
$r_{m}\simeq r_{c}$ to $r_{m}=1.5r_{c}$ and hence a small range of accretion
rates, as we observed for the 1 Hz QPO (see \S~\ref{fastdecayjump}).

The Spruit-Taam instability could modulate the accretion flow at high
amplitude, which would fit the observations of very high fractional
rms amplitudes for the QPO. The instability would also be compatible
with the continued presence of accretion-powered pulsations, since
accretion could still be funneled even if the inner edge of the disk
were oscillating.  Finally, the frequency of the instability has a
weak dependence on the mass accretion rate (see Fig. 4 in
\citealt{spr93}), rising or falling whether $r_{m}$ is greater or less
than $r_{c}$. This weak dependence has been observed in J1808
(Fig.~\ref{rms-flux-freq}) with the frequency rising with X-ray flux,
thus suggesting $r_{m}>r_{c}$.

\subsection{The mechanism for the 1 Hz QPO}

In summary, most of the mechanisms examined cannot, based on our
current understanding of how they work, explain key features of the 1
Hz QPO (see Table 2). The mechanisms that remain plausible are all
associated with, or fine-tuned by, the onset of the propeller regime.
%\begin{table*}
%%\centering
%\scriptsize
%\caption{Mechanisms for the 1 Hz QPO}
%\begin{tabular}{lcccccc}
%\hline
%\hline

\begin{deluxetable*}{lcccccc}
\tabletypesize{\footnotesize}
\tablecolumns{7}
%\tablewidth{0pt}
\tablecaption{Mechanisms for the 1 Hz QPO}
\tablehead{
  \colhead{Model} &
  \colhead{Freq.} &
  \colhead{Amp.} &
  \colhead{Amp./Energy} &
  \colhead{Flux threshold} &
  \colhead{Freq./Flux.}&
  \colhead{Propeller}
}
%MODEL & QPO Freq. & rms amplitude &  Energy  & flux treshold & freq.vs.flux & Propeller required ($r_m \sim r_c$)\\
%\hline
\startdata
& & & & & & \\
Surface oscillations  & {\bf P} & {\bf N} & {\bf Y} & {\bf N} &
\nodata  & {\bf N} \\

Marginally stable nuclear burning & {\bf N}  & \nodata & \nodata &
{\bf Y}  & \nodata & {\bf N}\\

Disk obscuration & {\bf Y}  & {\bf N} & {\bf P}  & \nodata & \nodata &
{\bf N}\\

Interchange instability ($r_m \ne r_c$) &  \nodata &  \nodata &
\nodata &  {\bf N} &
\nodata & {\bf N}\\

Interchange instability ($r_m \sim r_c$) & \nodata  &  \nodata &
\nodata &  {\bf Y} &
\nodata & {\bf Y}\\

Magnetic reconnection instability & {\bf P} & \nodata & \nodata  &
{\bf N}  & {\bf Y} & {\bf N} \\

Thermal/viscous instability & {\bf P} & \nodata & \nodata & {\bf
  N} &  \nodata & {\bf P}\\

Ionization instability & {\bf P} & \nodata & \nodata & {\bf Y} &
\nodata & {\bf P}\\

Radiation instability & {\bf N} & \nodata & \nodata & {\bf
  N} &  \nodata & {\bf N}\\

Spruit-Taam instability & {\bf P} & {\bf Y} & \nodata & {\bf Y} & {\bf Y} & {\bf Y}\\
\enddata
%\hline
%\hline
\tablecomments{The table compares observed properties of the 1 Hz QPO
   to the various models discussed in \S 4.  {\bf Y}/{\bf N} (yes/no)
  indicates that the model can/cannot explain the property in
  J1808 (e.g. at the accretion rates inferred for this source).  The symbol {\bf P} (possible) indicates that the model might be able to
  explain the property if certain conditions are met.  An empty field
  indicates that the model makes  no specific
prediction for that property, and that further studies are
required. Columns: (1) model, (2) 1 Hz QPO
frequency, (3) high fractional amplitude, (4) energy
dependence of the QPO fractional amplitude, (5) appearance of the
oscillation \textit{below} a certain flux threshold, (6) dependence of the
centroid frequency on the X-ray flux.  The last column shows
whether the onset of the propeller regime is necessary for the model
to be able to explain the observed QPO properties.  Only models
without {\bf N} in any column remain viable, pending further study.}
\end{deluxetable*}

There are a number of other pieces of evidence
(cf.~\S~\ref{1Hz:intro}) that also point to major changes in the
accretion environment at the luminosity where the 1 Hz QPO sets in
(\citealt{wij01},~\citealt{wij03},~\citealt{cam08}) - changes which
might be explained by the onset of the propeller. In addition there
are timing results suggesting a major change in disk structure around
this time, such as the $\sim0.2$ phase drift in the fundamental
(arguing for a major change in the disk environment around this time),
the change in the soft lag behavior \citep{har09b}, and the (debated)
detection of an accretion torque \citep{bur06, har08}.

The mechanism proposed by \citet{spr93} seems to be the most promising
candidate to explain the 1 Hz QPO, although the precise details of the
time scales for this instability in the situation when funnel flows
are relevant remain to be worked out.  It has a precise onset point
associated with the early propeller regime, should remain relatively
stable in frequency as accretion rate varies slightly and is only
expected in a narrow range of accretion rates.

Other mechanisms may also play a role, perhaps in concert with the
Spruit-Taam instability.  In \S~\ref{interchange} we mentioned that
new classes of interchange instabilities might operate near the
propeller transition, perhaps leading to sporadic accretion. In
\S~\ref{thermalvisc} we discussed the possibility of the ionization
instability triggering on short lengthscales in the inner regions of
the disk once the source enters the propeller regime.  This
possibility is particularly plausible if the disk is already close to
the transition from outburst to quiescence.  The ionization
instability might reinforce the Spruit-Taam instability mechanism, and
could also fine-tune the onset conditions for the 1 Hz QPO (see
\S~\ref{1808etal}).  The number of empty fields in Table 2 reflects
the scale of the modeling work required to resolve these questions.  

It is hard to understand why the 1 Hz QPO does not appear during the
faint re-flares in the 2008 outburst.  8 out of 57 observations were
in the 2-15 mCrab range during the re-flaring state.  The reason why
the 1 Hz QPO is not observed in these 8 observations is an open
problem. Although poorly constrained, the 1998 outburst exhibited a
similar behavior, and on several occasions during the 2000, 2002 and
2005 outbursts the 1 Hz QPO also remained undetected even for fluxes in
the 2-15 mCrab range, with fractional rms amplitude upper limits of
$\sim 10\%$. Clearly the 1 Hz QPO mechanism is not always triggered
even in the 2--15 mCrab range in J1808.

\subsection{J1808 and the other AMXPs}\label{1808etal}

\begin{figure*}[!th]
  \begin{center}
    \rotatebox{-90}{\includegraphics[width=1.0\columnwidth]{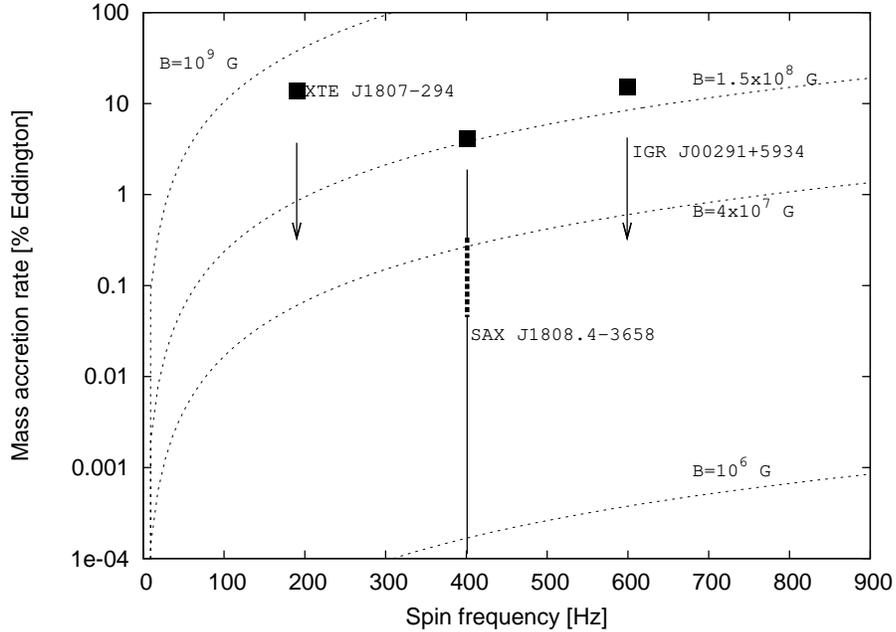}}
  \end{center}
  \caption{The accretion rate at which the propeller sets in for
  particular combinations of magnetic field and spin rate for the
  three AMXPs (dashed curves).  The vertical arrows show the range of
  accretion rates over which the AMXPs have been observed
  and refer to \emph{X-ray fluxes}. They have to be
  considered lower limits on the true mass accretion rates which are set
  by bolometric fluxes. The estimated \emph{bolometric fluxes} at
  the peak of the outburst (where the mass accretion should be the
  highest) are marked with a black square (see text for details). The
  thick dashed vertical line in J1808 indicates the range of mass
  accretion rates at which the 1 Hz QPO has been observed.
    \label{propeller}}
\end{figure*}

In order to enter the propeller regime, J1808 needs to be at the point
where $r_m \sim r_c$.  Equating the crude expressions given in
eq.(\ref{rc}) and (\ref{rm}), we obtain a relation between accretion
rate, magnetic field and spin rate.
Figure~\ref{propeller} shows the conditions for propeller onset for
particular combination of these parameters. 
Clearly this is very approximate, since it is based on the
simplest estimates of $r_m$ and $r_c$, and ignores dependencies on
mass and radius, but sufficient to understand whether the propeller
scenario is a realistic possibility.

We plot the mass accretion rate values of three well known AMXPs: XTE
J1807-294 (spin frequency 190 Hz), J1808 (401 Hz) and IGR J00291+294
(599 Hz).  The first object was chosen because its spin frequency is one
of the lowest known among AMXPs and its outburst spans a wide range of
luminosities.  IGR J00291+294 was chosen because its neutron star has
the highest spin frequency known among AMXPs.

The mass accretion rates used in Fig.~\ref{propeller} are calculated
for X-ray fluxes in the 2--10 keV energy band, by assuming a neutron
star mass of 1.4 $M_{\odot}$ and an efficiency of 10\% for the
conversion of rest mass energy of the accreted material into X-ray
flux.  Since these mass accretion rates do not refer to bolometric
fluxes, they have to be considered lower limits.  We also marked the
bolometric \emph{peak} luminosity of each source (as reported in
\citealt{gie02}, \citealt{fal05a, fal05b}) for assumed distances of
8.5 kpc (IGR J00291+294\footnote{we note that IGR J00291+294 is not
located toward the Galactic Center, therefore the assumed distance is
arbitrary, and is made for an easier comparison with XTE J1807-294}
and XTE J1807-294) and 3.5 kpc (J1808).  The very broad range of
luminosities of J1808 are observed thanks to the deeper observations
of {{\it Swift}-XRT} (\S~\ref{re-flarings}, \citealt{cam08}) and {{\it
XMM-Newton}} \citep{wij03}.

For all three sources, the conditions for propeller onset should be
encountered if the field strength is $\sim 10^{8}$ G.  For J1808, with
a spin of 401 Hz and an accretion rate that runs from a few percent of
Eddington at peak, down to less than 0.001 \% in the dips between the
re-flares, the system must always enter the propeller regime 
at same accretion rate, while for $B<10^{6}$ G the system will not enter
the propeller regime in the observed range of mass accretion rates.  The
range of magnetic fields is ($B\sim 0.4$--$1.5\times 10^{8}$G) as
reported by \citet{har08, har09}.  The range of accretion rates for
which the 1 Hz QPO appears (inferred from the 2-15 mCrab X-ray flux,
\S~\ref{fastdecayjump}) lies just below this range. This coincidence
is quite impressive since the accretion rates are lower limits.

By looking at Fig.~\ref{propeller} we can infer that the main reason
why the 1 Hz QPO has been observed in J1808 and not in other AMXPs
might be related with the proximity of J1808 (3.5 kpc) with respect to
the other AMXPs (assuming they are all located at a distance close to
$8.5$kpc).
From a visual inspection of Fig.~\ref{propeller} the non-observation
of the propeller in IGR J00291 could be problematic if the maximum
accretion rate is $\sim10\%$ Eddington and the magnetic field is
$\sim10^{8}$G.  However, many uncertainties can play a role here:
uncertainities on the source distance, the conversion of X-ray
luminosities into mass accretion rates, the precise condition for the
onset of the propeller, not to mention all the uncertainties related
with the definition of magnetospheric radius. In this sense
Fig.~\ref{propeller} has to be taken as a qualitative picture, useful
to understand the underlying behavior of this sources, but with too
many uncertainties remaining to draw robust conclusions.

However, if we suppose that the instability is triggered in IGR
J00291+5934 at the same mass accretion rates as in J1808, then the
expected luminosity would be below the detection threshold of {{\it
RXTE}}. None of the known AMXPs has been extensively monitored at low
flux levels by {{\it Swift}-XRT} or {{\it XMM-Newton}}, both of which
would be able to (easily) test this scenario.

If J1808 enters the strong propeller regime during the re-flares, a
large outflow of gas is expected.  \citet{har08} and \citet{dis08}
observed an anomalously large orbital period derivative. A possible
explanation requires a mass loss from the system of $\sim 10^{-9}$M$_{\odot}$
yr$^{-1}$ (\citealt{dis08} and \citealt{bur09} for a description of
this scenario). The onset of a strong propeller with outflows of
matter from the system can in principle play a role in this.
If however, the real explanation is different (see \citealt{har09} for
a discussion of alternative possibilities to the strong outflow 
scenario) then the propeller might have only a minor role, if any, 
on the long term evolution of the orbital period. 

A strong propeller might in principle explain the long term spin
down of the neutron star in J1808 as proposed by \citet{har08}.  By
assuming a B field of $\sim10^8$G, and assuming $r_{m}\sim 2\,r_{c}$
we can calculate the amount of mass that needs to be ejected in a
strong propeller regime to produce the long term spin down observed by
Hartman et al. (2009a) ($\dot{\nu}\sim-5.5\times10^{-16}\rm\,Hz\,s^{-1}$):
\begin{equation}
N_{sd} =N_{prop}\rightarrow  2\pi\,I\dot{\nu}=\sqrt{GMr_{m}}\dot{M}_{ej} 
\end{equation}
where $N_{sd}$ is the spin-down torque, $N_{prop}$ is the propeller
torque (defined as in \citealt{bil98b}) and $\dot{M}_{ej}$ is the amount of mass expelled during the
strong propeller stage. The mass ejection required during quiescence
is $\sim 8\times 10^{-11}\rm\,M_{\odot}\,yr^{-1}$. 
This mass outflow is still insufficient to explain
the large orbital period derivative observed in J1808.
Moreover, even if the disk material is expelled at $r_{m}\sim 2\,r_{c}$, 
the viscous dissipation of gravitational potential energy 
would yield a luminosity of $\approx 10^{35}\rm\,erg\,s^{-1}$, which
is too high when compared to the quiescence luminosity of J1808
($\sim 5\times 10^{31}\rm\,erg\,s^{-1}$, \citealt{cam02}, \citealt{hei09}).

We have shown that the 1 Hz modulation has an effect on the 401 Hz
pulse formation. Magnetic channeling is expected to be easier in the
valleys than in the peaks of the 1 Hz modulation, where the accretion
rate is lower. We observed an increase of the 401 Hz pulsed fraction
in the valleys in agreement with this. We have also found possible
links between the 1 Hz modulation amplitude and the 401 Hz pulse phase.
If the 1 Hz QPO reflects a change in the accretion flow, the position
of the magnetic funnel can change accordingly, thus affecting the
401 Hz pulse phase. A better understanding of the process
that generates the 1 Hz modulation would be fundamental to clarify
this magnetosphere/disk interaction.

\section{Conclusions}

We have performed the first complete study of the 1 Hz modulation,
its relation to the 401 Hz
pulsations and the re-flares of SAX J1808.4-3658 as observed over 10 years.
Several common features are observed in the 2000, 2002 and 2005
outbursts while the 1998 and 2008 outbursts have different properties,
the most remarkable one being the absence of a strong 1 Hz modulation.

We focused on the origin of the 1 Hz oscillation that sometimes
dominates the re-flare lightcurve and we found that all viable
candidate mechanisms are connected with the onset of the propeller
stage. The most promising model discussed is the Spruit-Taam
instability which explains the stable 1 Hz frequency, its high
amplitude and the narrow flux range of its occurrence in a natural
way, and is also compatible with the simultaneous presence of 401 Hz
pulsations.

Many open issues remain. It is not
clear yet why the pulse phase drifts by 0.2 cycles during the fast
decay just before the onset of the 1 Hz modulation. It is also unclear
whether the pulse phase jumps observed on time scales of hours or less are
related with the amplitude of the 1 Hz. However, it is likely that this
pulse behavior reflects changes in the accretion flow onto the surface
triggered by flow changes associated with the 1 Hz QPO mechanism.

The reason why 1 Hz modulation has not been observed in other AMXPs
might be the larger distance of the other sources making them
undetectable in the relevant luminosity range. Future monitoring of
low level flux states of other AMXPs will be very important to extend our
comprehension of the 1 Hz modulation and further probe the onset of
the propeller stage.

\acknowledgements{Acknowledgements: We would like to thank the {\it
Swift} team for promptly scheduling the 2008 observations of SAX
J1808.4-3658. We thank also P. Casella and D. Altamirano for useful
discussions. We would also like to thank Henk Spruit for comments on
an earlier version of this manuscript.}

\clearpage
\end{document}